\documentclass[aps,pra,twocolumn,groupedaddress,superscriptaddress]{revtex4}

\usepackage[dvips]{graphicx}

\begin{document}


\title{Inducing vortices in a Bose-Einstein condensate using \\
holographically produced light beams}



\author{J. F. S. Brachmann}
\affiliation{MIT-Harvard Center for Ultracold Atoms, Cambridge, MA 02138, USA}
\affiliation{Physics Department, Harvard University, Cambridge, MA 02138, USA}
\affiliation{Ludwig-Maximilians-Universit\"at, Schellingstra\ss e 4, 80799 M\"unchen, Germany}

\author{W. S. Bakr}
\affiliation{MIT-Harvard Center for Ultracold Atoms, Cambridge, MA 02138, USA}
\affiliation{Physics Department, Harvard University, Cambridge, MA 02138, USA}

\author{J. Gillen}
\affiliation{MIT-Harvard Center for Ultracold Atoms, Cambridge, MA 02138, USA}
\affiliation{Physics Department, Harvard University, Cambridge, MA 02138, USA}
\affiliation{Physics Department, Massachusetts Institute of Technology, Cambridge, MA 02139, USA}

\author{A. Peng}
\affiliation{MIT-Harvard Center for Ultracold Atoms, Cambridge, MA 02138, USA}
\affiliation{Physics Department, Harvard University, Cambridge, MA 02138, USA}
\affiliation{Washington University School of Medicine, Saint Louis, MO 63110, USA}

\author{M. Greiner}
\affiliation{MIT-Harvard Center for Ultracold Atoms, Cambridge, MA 02138, USA}
\affiliation{Physics Department, Harvard University, Cambridge, MA 02138, USA}


\date{\today}

\begin{abstract}
In this paper we demonstrate a technique that can create out-of-equilibrium vortex configurations with almost arbitrary charge and geometry in a Bose-Einstein condensate. We coherently transfer orbital angular momentum from a holographically generated light beam to a $^{87}$Rb condensate using a two-photon stimulated Raman process. Using matter wave interferometry, we verify the phase pattern imprinted onto the atomic wave function for a single vortex and a vortex-antivortex pair. In addition to their phase winding, the vortices created with this technique have an associated hyperfine spin texture.
\end{abstract}


\maketitle


\par Light fields that carry orbital angular momentum (OAM), have attracted a lot of attention from researchers in fields as diverse as biophysics \cite{grier:optical_manipulation}, micromechanics and -fabrication \cite{galajda:micromachines_driven_by_light}, astronomy \cite{gelly:aom_in_astronomical_objects}, quantum communication \cite{zeilinger:oam_photon_entanglement} and the field of ultracold atoms \cite{molina:twisted_photons}. While spin angular momentum is associated with the polarization of the light field, orbital angular momentum is associated with the spatial mode of the field. In the paraxial approximation, both angular momentum components can be measured separately \cite{enk}. It has been shown \cite{allen} that paraxial Laguerre-Gaussian laser beams, which are feasible to create experimentally, carry a well-defined orbital angular momentum associated with their spiral wavefronts. More generally, light beams that contain a phase singularity, carry OAM parallel or anti-parallel to their propagation direction depending on the phase variation around the singularity. If more than one singularity is present, $(q_{left}-q_{right}) \hbar$ OAM quanta are contained in the beam. Here the quantum numbers $q_{left}$ and $q_{right}$ determine the charge and chirality (left- or right handed spiraling of the wavefront) of the optical vortices. 
\par The torque produced through momentum exchange between an OAM light beam and matter can be used to induce rotation of the so illuminated object. In biophysics and micromechanics this torque has been proposed to be utilizable in driving molecular motors and micromachines, which can at the same time be trapped at the location of the singularity, where the beam intensity has to vanish \cite{grier:optical_manipulation,petrov:torque_detection,Rubinsztein-Dunlop:transfer_OAM_particles,Dholakia:OAM_transfer_optically_trapped_particle}. Also, in the field of ultracold atoms, direct transfer of OAM from light to atoms in an ultracold gas cloud offers interesting possibilities \cite{phillips:orbital_angular_momentum_light_beams_quantized_vortices, bigelow:zeemanstates_oam_beam_bec_vortex, bigelow:sculpture_spinor_bec, bigelow:skyrmions}.
\par An analogue of optical vortices can be realized in superfluids, where orbital angular momentum can only be added in the form of quantized vortices. In liquid helium, vortices have been studied experimentally since the 1960s \cite{reif:superfluid_helium_vortices}. With the achievement of Bose-Einstein condensation (BEC) in dilute atomic gases, a very controllable macroscopic quantum object is available, which has already been used extensively for further study of quantized vortex states\cite{ketterle:bec_vortex_lattice, ketterle:dynamical_instability_doubly_quantized_vortex, cornell:vortex_precession_filled_and_empty_cores, hall:dynamics_vortex_lines}. In the regime where the number of vortices is much smaller than the number of atoms in the condensate, the manybody wavefunction of the BEC can be approximated using mean field theory. In this description, every particle in the condensate carries the same angular momentum, quantized in units of $\hbar$.
\par Bose-Einstein condensates have opened the path to studying many new regimes in vortex physics. For example, because of the analogy to quantum Hall physics, the low filling factor regime of atoms to vortices is interesting to reach \cite{cornell:lowest_landau_level, sorensen:quantum_hall,2010arXiv1007}. Also, the  Berezinskii-Kosterlitz-Thouless phase transition characterized by the binding/unbinding of thermally activated vortex-antivortex pairs in two-dimensional gases is subject of current research \cite{kosterlitz:bkt, dalibard:2D_vortices}.

\par Although the first vortices in BECs were produced by a phase imprinting technique \cite{cornell:vortices}, vortices are most commonly produced by stirring with an anisotropic trapping potential. This can be created for example by the use of a rotating detuned laser beam \cite{ketterle:bec_vortex_lattice}. In this way, Abrikosov lattices of vortices with the same circulation direction are created. With this technique, it is not easy to create vortices with different circulation directions, non-equilibrium vortex states such as multiply charged vortices, arbitrary vortex lattices, or superpositions of states with different vorticity. However, other methods of producing vortices have demonstrated some of these possibilities \cite{matthews:vortices, leanhardt:topological_vortices,PhysRevLett.104.160401}.
\par Recently, a different approach to create vortices has been demonstrated: Orbital angular momentum was transfered in units of $\hbar$ from photons in a Laguerre-Gaussian beam to atoms in a BEC using a two-photon Raman process \cite{phillips:orbital_angular_momentum_light_beams_quantized_vortices, bigelow:zeemanstates_oam_beam_bec_vortex, bigelow:sculpture_spinor_bec, bigelow:skyrmions}. In this way, superpositions of vortex states could be created in a controlled manner.
\par
\begin{figure}[t]
\includegraphics[width=3.3in]{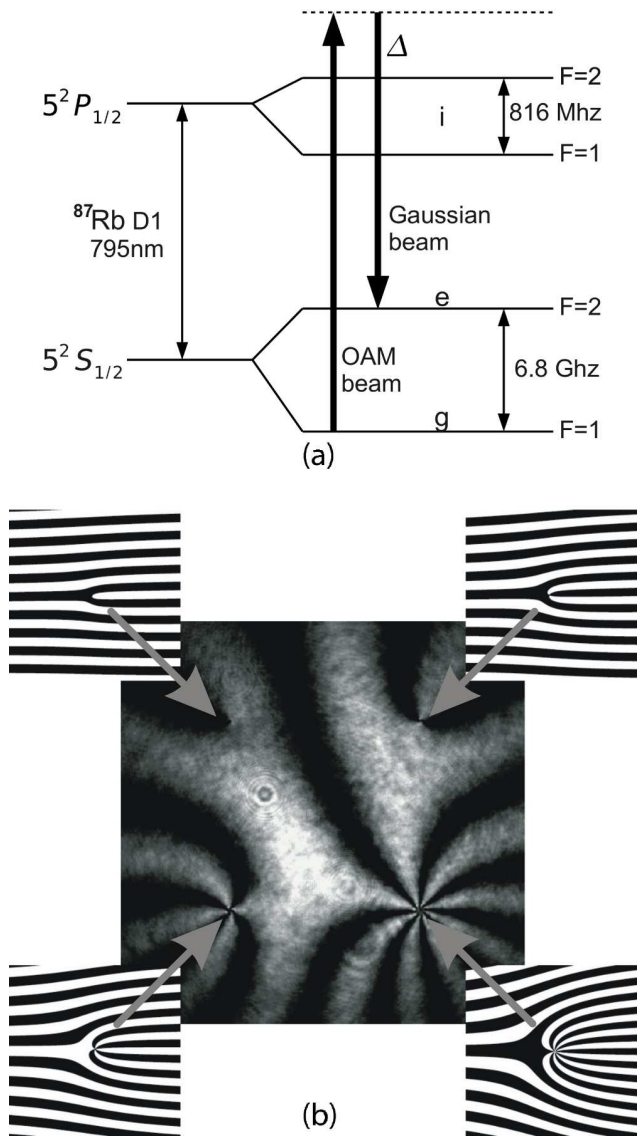}
\caption{(a) Raman scheme for $^{87}$Rb, D1 line, using a Gaussian beam and the first order diffracted beam from a hologram. $\Delta$ is the common detuning of both beams from the intermediate $\vert 5^{2}P_{1/2}; F=2 \rangle$ state. (b) interference pattern of a Gaussian beam with the first order diffracted beam from a hologram for producing optical vortices with charges 1,2,4 and 8. The key structures from the original binary hologram producing the respective phase windings are shown as cut outs and are related to the interference pattern.}
\label{ramansketch}
\end{figure}

 To be able to really take advantage of this method, one needs the ability to create OAM beams in a versatile way. Since a hologram is, in the paraxial approximation, capable of creating any desired wavefront, it can be the right tool for OAM beam creation if manufactured in good enough quality. We make use of standard lithographic techniques for chrome mask manufacturing to obtain binary approximations of greyscale holograms for different OAM beams in very good quality. Figure \ref{ramansketch}(b) shows cut-outs of a hologram used to produce multiply charged optical vortices as an example. 
\par In this paper we present results from inducing vortices by transferring OAM from a holographically produced light beam to a $^{87}$Rb BEC using a two-photon Raman process shown in Figure \ref{ramansketch}(a). With a suitable hologram, we are able to directly produce a rotational state of the BEC with a vortex-antivortex pair~\cite{PhysRevLett.104.160401}. The coherence of the process is shown by atom interferometry, with which it can be seen that the phase of the holographically produced light beam is imprinted onto the condensate wavefunction.
\par Our holograms do not create Laguerre-Gaussian beams, but rather beams with a Gaussian profile except at the points of the phase singularities. The size of the optical vortices is determined by the resolution of the projection optics (about $5 \mu$m). By using Gaussian beams with a large waist, the intensity profile away from the vortex core becomes uniform over the condensate, ensuring a homogeneous transfer rate to the vortex state. Since the beam intensity close to the phase singularity is nevertheless close to zero, atoms in the vortex core do not undergo the Raman transition. As a result, the vortices that we produce have cores filled with a different spin state. Both spin states can be imaged separately.
\par In the experiment, the atoms start in the $F=1$ state and are coherently transferred to the $F=2$ state by illuminating them with two copropagating beams, a Gaussian beam and a beam carrying OAM that is produced by a hologram (Figure \ref{ramansketch}(a)). An atom transferring a photon from the hologram beam to the Gaussian beam gains $q \hbar$ orbital angular momentum quanta, where $q$ is the number of OAM quanta in the hologram beam. Since the beams are copropagating, no net linear momentum is transferred to the atoms by this process.
\par The difference in frequency of the two light fields is close to $6.8 $GHz corresponding to the hyperfine splitting between the initial $F=1$ and final $F=2$ state of the transition. $\sigma^{+}$ - $\sigma^{-}$ polarizations are chosen, resulting in a change of the magnetic state by $\Delta m_F = + 2$. A weak homogeneous magnetic field pointing in the same direction as the beams is applied during the Raman pulses for the definition of a quantization axis.
\par Our experiments begin with a $^{87}$Rb BEC of approximately $3 \times 10^5$ atoms in the $\vert 5^{2}S_{1/2}; F=1, m_F=-1 \rangle$ state. The condensate is in a spherical quadrupole-Ioffe configuration (QUIC) trap with a trapping frequency of $\omega = 2\pi \times 20 Hz$ and has a Thomas-Fermi radius of approximately $15 \mu m$. $14 ms$ after releasing the atoms from the magnetic trap, when the radius of the cloud is about $30 \mu m$, we pulse on the two Raman beams to project the vortex pattern onto the expanded cloud.  The beams, travelling horizontally, are applied as square pulses to the vertically falling cloud and change the internal atomic state to $\vert 5^{2}S_{1/2}; F=2, m_F=1 \rangle$. Beam powers of $10 \mu W$ and waists of $131 \mu m$ are used with square pulse durations of $8 \mu s$. The common detuning of both Raman beams from the $\vert 5^{2}P_{1/2}; F=2 \rangle$ state is set to $\Delta = 4.18$GHz. Spontaneous emission during the Raman pulse duration is negligible and does not introduce decoherence relevant to the experiment. Absorption images of the condensate are taken $3 ms$ after applying the Raman pulses.
\begin{figure}[b]
\includegraphics[width=2.7in]{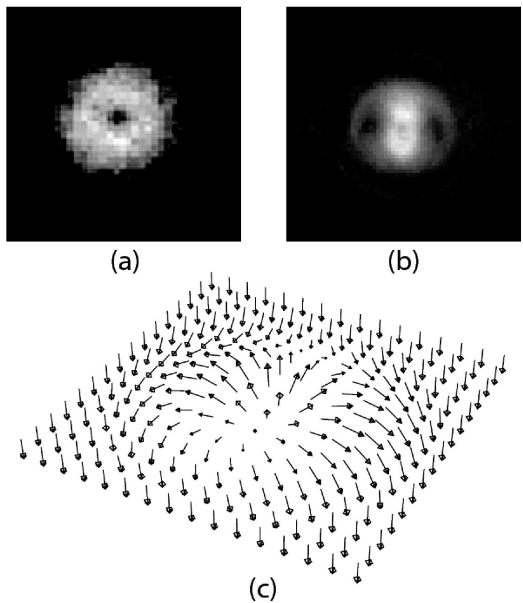}
\caption{(a) Absorption image of a condensate with a single vortex and (b) two counter-rotating vortices. (c) A spin texture similar to a skyrmion corresponding to the core region of the vortex in (a).}
\label{single_and_double_vortex}
\end{figure}
\par Figures \ref{single_and_double_vortex}(a),(b) show absorption images of BECs with a single vortex and two counter-rotating vortices respectively. In the pictures, only atoms projected onto $\vert 5^{2}S_{1/2}; F=2, m_F=1 \rangle$ are imaged, although a fraction of the condensate in $\vert 5^{2}S_{1/2}; F=1, m_F=-1 \rangle$ is still present. The vortex core sizes in these pictures is limited by the resolution of the projection optics.
\par We emphasize here that this technique produces spin vortices~\cite{PhysRevLett.103.080603}, as opposed to stirring techniques which produce charge (density) vortices. To produce the vortex in Figure \ref{single_and_double_vortex}(a), the intensity/pulse length is adjusted to drive a $\pi$ Raman transition far enough from the singularity. If we regard $F=1,2$ as a two level system, the spin vector rotates from ($F=1$) at the center of the vortex to ($F=2$) at the edge of the core region, where the OAM beam intensity becomes constant. At the same time, the projection of the spin vector in a plane perpendicular to the direction of the beam
propagation rotates azimuthally by $2 \pi$ due to the phase winding in the optical vortex. This spin texture, shown in Figure \ref{single_and_double_vortex}(c), describes a skyrmion. Skrymions have been used as a model for baryonic particles in nuclear physics \cite{khawaja:skyrmions_in_ferromagnetic_bec,skyrme:nonlinear_field_theory}. Multi-component vortices have also been studied previously in BECs by other authors \cite{ketterle:coreless_vortex_bec, matthews:vortices, bigelow:sculpture_spinor_bec, bigelow:skyrmions}. 
\par
\begin{figure}[t]
\includegraphics[width=3.1in]{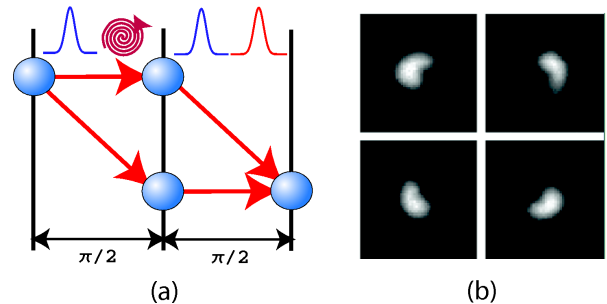}
\caption{(a) Pulse sequence for atom interferometry with two Raman pulses (details in text) (b) Absorption images for different phases of a vortex state of the BEC interfering with a uniform phase state.}
\label{interferometry}
\end{figure}

 To show the coherence of the process, the atomic vortex state is interfered with an atomic plane uniform phase state. This is done by applying two separate $\frac{\pi}{2}$ Raman pulses as shown in Figure \ref{interferometry}(a): the first pulse uses the hologram beam and the Gaussian beam to transfer a fraction of atoms into a vortex state as before. Immediately afterwards, a second pulse is executed where the hologram beam is replaced with a second copropagating Gaussian beam at the same frequency to transfer another fraction of atoms to the same internal state, this time with a uniform phase. The pulse lengths used are $4\mu s$  each.
\par Figure \ref{interferometry}(b) shows absorption images for different relative phases of the wavefunctions of the vortex state interfering with the uniform phase state. The phase difference between the hologram beam and the phase reference Gaussian beam maps directly onto the atoms.
\par
\begin{figure}[b]
\includegraphics[width=3.3in]{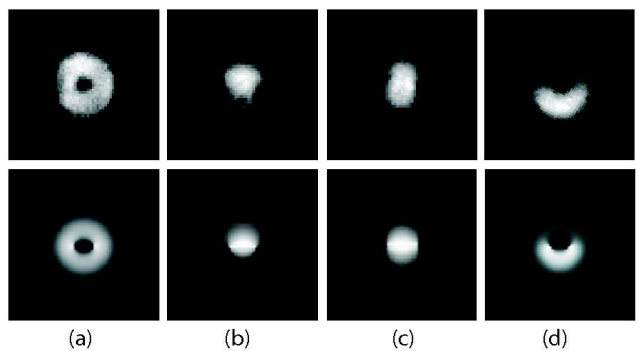}
\caption{Top row: absorption images for a vortex-antivortex state interfering with a uniform phase state. Bottom row: calculated interference patterns. The results are shown for phase differences of (a) $0\pi/5$, (b) $4\pi/5$, (c) $5\pi/5$ and (d) $9\pi/5$. The phases of the vortex state in the simulated data where chosen to match the experimental data.}
\label{vortex_antivortex_interferometry}
\end{figure}

 In Figure \ref{vortex_antivortex_interferometry}, we show the results of atom interferometry on a BEC state with a vortex-antivortex pair, together with simulated data. From these images, we verify that the two vortices are counter-rotating, as their phases wind in opposite directions as the phase between the light beams is varied.
\par
\begin{figure}[t]
\includegraphics[width=2.2in]{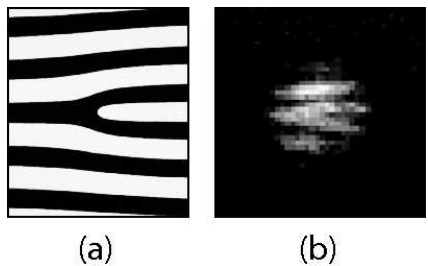}
\caption{A matter wave hologram (b) resembling the original hologram (a) (microscopy image of the chrome mask) is created if linear momentum is transferred in the second pulse of the described sequence.}
\label{tilted_beams}
\end{figure}

 If the hologram beam and the reference Gaussian beam used for the interference experiments are not perfectly copropogating, the structure of the original hologram can be reproduced in the atomic cloud as an interference between the two states. Figure \ref{tilted_beams} shows an interference image for a single vortex state obtained in this way. The relative linear momentum between the two interfering condensates appears as a sinusoidally varying pattern of stripes on top of the interference pattern due to the vortex \cite{ketterle:interference_2_becs}.
\par In this paper, we have demonstrated the controlled coherent preparation of vortex states in Bose-Einstein condensates. The phase of the classical electromagnetic field, which can be shaped holographically in a versatile way, is imprinted onto the atomic cloud, allowing great freedom in the phase pattern to be transferred. Vortex patterns with arbitrary charge, circulation and configurations can be created.
\par We have recently implemented high resolution optics in our apparatus ($\sim  600$nm resolution)\cite{greiner:quantum_gas_microscope}. The projection and imaging of more complex vortex patterns is now within reach. The high resolution should also enable reaching low filling factors of atoms to vortices, opening an avenue to studies in the quantum Hall regime. Finally, \textit{in-situ} projection of vortex patterns would allow studying out-of-equilibrium dynamics of vortex states and addressing questions about the stability of skyrmions, merons and other topological spin excitations. 

\par This work was supported by grants from the Army Research Office with funding from the DARPA OLE program, an AFOSR MURI program, and by grants from the NSF.


\begin{thebibliography}{34}
\expandafter\ifx\csname natexlab\endcsname\relax\def\natexlab#1{#1}\fi
\expandafter\ifx\csname bibnamefont\endcsname\relax
  \def\bibnamefont#1{#1}\fi
\expandafter\ifx\csname bibfnamefont\endcsname\relax
  \def\bibfnamefont#1{#1}\fi
\expandafter\ifx\csname citenamefont\endcsname\relax
  \def\citenamefont#1{#1}\fi
\expandafter\ifx\csname url\endcsname\relax
  \def\url#1{\texttt{#1}}\fi
\expandafter\ifx\csname urlprefix\endcsname\relax\def\urlprefix{URL }\fi
\providecommand{\bibinfo}[2]{#2}
\providecommand{\eprint}[2][]{\url{#2}}

\bibitem[{\citenamefont{Grier}(2003)}]{grier:optical_manipulation}
\bibinfo{author}{\bibfnamefont{D.~G.} \bibnamefont{Grier}},
  \bibinfo{journal}{Nature} \textbf{\bibinfo{volume}{424}},
  \bibinfo{pages}{810} (\bibinfo{year}{2003}).

\bibitem[{\citenamefont{Galajda and
  Ormos}(2001)}]{galajda:micromachines_driven_by_light}
\bibinfo{author}{\bibfnamefont{P.}~\bibnamefont{Galajda}} \bibnamefont{and}
  \bibinfo{author}{\bibfnamefont{P.}~\bibnamefont{Ormos}},
  \bibinfo{journal}{Appl. Phys. Lett.} \textbf{\bibinfo{volume}{78}},
  \bibinfo{pages}{249} (\bibinfo{year}{2001}).

\bibitem[{\citenamefont{{Uribe-Patarroyo, N.}
  et~al.}(2011)\citenamefont{{Uribe-Patarroyo, N.}, {Alvarez-Herrero, A.},
  {L\'opez Ariste, A.}, {Asensio Ramos, A.}, {Belenguer, T.}, {Manso Sainz,
  R.}, {LeMen, C.}, and {Gelly, B.}}}]{gelly:aom_in_astronomical_objects}
\bibinfo{author}{\bibnamefont{{Uribe-Patarroyo, N.}}},
  \bibinfo{author}{\bibnamefont{{Alvarez-Herrero, A.}}},
  \bibinfo{author}{\bibnamefont{{L\'opez Ariste, A.}}},
  \bibinfo{author}{\bibnamefont{{Asensio Ramos, A.}}},
  \bibinfo{author}{\bibnamefont{{Belenguer, T.}}},
  \bibinfo{author}{\bibnamefont{{Manso Sainz, R.}}},
  \bibinfo{author}{\bibnamefont{{LeMen, C.}}}, \bibnamefont{and}
  \bibinfo{author}{\bibnamefont{{Gelly, B.}}}, \bibinfo{journal}{A\&A}
  \textbf{\bibinfo{volume}{526}}, \bibinfo{pages}{A56} (\bibinfo{year}{2011}).

\bibitem[{\citenamefont{Mair et~al.}(2001)\citenamefont{Mair, Vaziri, Weihs,
  and Zeilinger}}]{zeilinger:oam_photon_entanglement}
\bibinfo{author}{\bibfnamefont{A.}~\bibnamefont{Mair}},
  \bibinfo{author}{\bibfnamefont{A.}~\bibnamefont{Vaziri}},
  \bibinfo{author}{\bibfnamefont{G.}~\bibnamefont{Weihs}}, \bibnamefont{and}
  \bibinfo{author}{\bibfnamefont{A.}~\bibnamefont{Zeilinger}},
  \bibinfo{journal}{Nature} \textbf{\bibinfo{volume}{412}},
  \bibinfo{pages}{313} (\bibinfo{year}{2001}).

\bibitem[{\citenamefont{Molina-Terriza
  et~al.}(2007)\citenamefont{Molina-Terriza, Torres, and
  Torner}}]{molina:twisted_photons}
\bibinfo{author}{\bibfnamefont{G.}~\bibnamefont{Molina-Terriza}},
  \bibinfo{author}{\bibfnamefont{J.~P.} \bibnamefont{Torres}},
  \bibnamefont{and} \bibinfo{author}{\bibfnamefont{L.}~\bibnamefont{Torner}},
  \bibinfo{journal}{Nature Phys.} \textbf{\bibinfo{volume}{3}},
  \bibinfo{pages}{305} (\bibinfo{year}{2007}).

\bibitem[{\citenamefont{van Enk and Nienhuis}(1994)}]{enk}
\bibinfo{author}{\bibfnamefont{S.~J.} \bibnamefont{van Enk}} \bibnamefont{and}
  \bibinfo{author}{\bibfnamefont{G.}~\bibnamefont{Nienhuis}},
  \bibinfo{journal}{Europhys. Lett.} \textbf{\bibinfo{volume}{25}},
  \bibinfo{pages}{497} (\bibinfo{year}{1994}).

\bibitem[{\citenamefont{Allen et~al.}(1992)\citenamefont{Allen, Beijersbergen,
  Spreeuw, and Woerdman}}]{allen}
\bibinfo{author}{\bibfnamefont{L.}~\bibnamefont{Allen}},
  \bibinfo{author}{\bibfnamefont{M.~W.} \bibnamefont{Beijersbergen}},
  \bibinfo{author}{\bibfnamefont{R.~J.~C.} \bibnamefont{Spreeuw}},
  \bibnamefont{and} \bibinfo{author}{\bibfnamefont{J.~P.}
  \bibnamefont{Woerdman}}, \bibinfo{journal}{Phys. Rev. A}
  \textbf{\bibinfo{volume}{45}}, \bibinfo{pages}{8185} (\bibinfo{year}{1992}).

\bibitem[{\citenamefont{Volpe and Petrov}(2006)}]{petrov:torque_detection}
\bibinfo{author}{\bibfnamefont{G.}~\bibnamefont{Volpe}} \bibnamefont{and}
  \bibinfo{author}{\bibfnamefont{D.}~\bibnamefont{Petrov}},
  \bibinfo{journal}{Phys. Rev. Lett.} \textbf{\bibinfo{volume}{97}},
  \bibinfo{pages}{210603} (\bibinfo{year}{2006}).

\bibitem[{\citenamefont{He et~al.}(1995)\citenamefont{He, Friese, Heckenberg,
  and Rubinsztein-Dunlop}}]{Rubinsztein-Dunlop:transfer_OAM_particles}
\bibinfo{author}{\bibfnamefont{H.}~\bibnamefont{He}},
  \bibinfo{author}{\bibfnamefont{M.~E.~J.} \bibnamefont{Friese}},
  \bibinfo{author}{\bibfnamefont{N.~R.} \bibnamefont{Heckenberg}},
  \bibnamefont{and}
  \bibinfo{author}{\bibfnamefont{H.}~\bibnamefont{Rubinsztein-Dunlop}},
  \bibinfo{journal}{Phys. Rev. Lett.} \textbf{\bibinfo{volume}{75}},
  \bibinfo{pages}{826} (\bibinfo{year}{1995}).

\bibitem[{\citenamefont{Garc\'es-Ch\'avez
  et~al.}(2003)\citenamefont{Garc\'es-Ch\'avez, McGloin, Padgett, Dultz,
  Schmitzer, and Dholakia}}]{Dholakia:OAM_transfer_optically_trapped_particle}
\bibinfo{author}{\bibfnamefont{V.}~\bibnamefont{Garc\'es-Ch\'avez}},
  \bibinfo{author}{\bibfnamefont{D.}~\bibnamefont{McGloin}},
  \bibinfo{author}{\bibfnamefont{M.~J.} \bibnamefont{Padgett}},
  \bibinfo{author}{\bibfnamefont{W.}~\bibnamefont{Dultz}},
  \bibinfo{author}{\bibfnamefont{H.}~\bibnamefont{Schmitzer}},
  \bibnamefont{and} \bibinfo{author}{\bibfnamefont{K.}~\bibnamefont{Dholakia}},
  \bibinfo{journal}{Phys. Rev. Lett.} \textbf{\bibinfo{volume}{91}},
  \bibinfo{pages}{093602} (\bibinfo{year}{2003}).

\bibitem[{\citenamefont{Andersen et~al.}(2006)\citenamefont{Andersen, Ryu,
  Clade, Natarajan, Vaziri, Helmerson, and
  Phillips}}]{phillips:orbital_angular_momentum_light_beams_quantized_vortices}
\bibinfo{author}{\bibfnamefont{M.~F.} \bibnamefont{Andersen}},
  \bibinfo{author}{\bibfnamefont{C.}~\bibnamefont{Ryu}},
  \bibinfo{author}{\bibfnamefont{P.}~\bibnamefont{Clade}},
  \bibinfo{author}{\bibfnamefont{V.}~\bibnamefont{Natarajan}},
  \bibinfo{author}{\bibfnamefont{A.}~\bibnamefont{Vaziri}},
  \bibinfo{author}{\bibfnamefont{K.}~\bibnamefont{Helmerson}},
  \bibnamefont{and} \bibinfo{author}{\bibfnamefont{W.~D.}
  \bibnamefont{Phillips}}, \bibinfo{journal}{Phys. Rev. Lett.}
  \textbf{\bibinfo{volume}{97}}, \bibinfo{eid}{170406} (\bibinfo{year}{2006}).

\bibitem[{\citenamefont{Wright et~al.}(2008)\citenamefont{Wright, Leslie, and
  Bigelow}}]{bigelow:zeemanstates_oam_beam_bec_vortex}
\bibinfo{author}{\bibfnamefont{K.~C.} \bibnamefont{Wright}},
  \bibinfo{author}{\bibfnamefont{L.~S.} \bibnamefont{Leslie}},
  \bibnamefont{and} \bibinfo{author}{\bibfnamefont{N.~P.}
  \bibnamefont{Bigelow}}, \bibinfo{journal}{Phys. Rev. A}
  \textbf{\bibinfo{volume}{77}}, \bibinfo{eid}{041601} (\bibinfo{year}{2008}).

\bibitem[{\citenamefont{Wright et~al.}(2009)\citenamefont{Wright, Leslie,
  Hansen, and Bigelow}}]{bigelow:sculpture_spinor_bec}
\bibinfo{author}{\bibfnamefont{K.~C.} \bibnamefont{Wright}},
  \bibinfo{author}{\bibfnamefont{L.~S.} \bibnamefont{Leslie}},
  \bibinfo{author}{\bibfnamefont{A.}~\bibnamefont{Hansen}}, \bibnamefont{and}
  \bibinfo{author}{\bibfnamefont{N.~P.} \bibnamefont{Bigelow}},
  \bibinfo{journal}{Phys. Rev. Lett.} \textbf{\bibinfo{volume}{102}},
  \bibinfo{pages}{030405} (\bibinfo{year}{2009}).

\bibitem[{\citenamefont{Leslie et~al.}(2009)\citenamefont{Leslie, Hansen,
  Wright, Deutsch, and Bigelow}}]{bigelow:skyrmions}
\bibinfo{author}{\bibfnamefont{L.~S.} \bibnamefont{Leslie}},
  \bibinfo{author}{\bibfnamefont{A.}~\bibnamefont{Hansen}},
  \bibinfo{author}{\bibfnamefont{K.~C.} \bibnamefont{Wright}},
  \bibinfo{author}{\bibfnamefont{B.~M.} \bibnamefont{Deutsch}},
  \bibnamefont{and} \bibinfo{author}{\bibfnamefont{N.~P.}
  \bibnamefont{Bigelow}}, \bibinfo{journal}{Phys. Rev. Lett.}
  \textbf{\bibinfo{volume}{103}}, \bibinfo{pages}{250401}
  (\bibinfo{year}{2009}).

\bibitem[{\citenamefont{Rayfield and
  Reif}(1963)}]{reif:superfluid_helium_vortices}
\bibinfo{author}{\bibfnamefont{G.~W.} \bibnamefont{Rayfield}} \bibnamefont{and}
  \bibinfo{author}{\bibfnamefont{F.}~\bibnamefont{Reif}},
  \bibinfo{journal}{Phys. Rev. Lett.} \textbf{\bibinfo{volume}{11}},
  \bibinfo{pages}{305} (\bibinfo{year}{1963}).

\bibitem[{\citenamefont{Abo-Shaeer et~al.}(2001)\citenamefont{Abo-Shaeer,
  Raman, Vogels, and Ketterle}}]{ketterle:bec_vortex_lattice}
\bibinfo{author}{\bibfnamefont{J.~R.} \bibnamefont{Abo-Shaeer}},
  \bibinfo{author}{\bibfnamefont{C.}~\bibnamefont{Raman}},
  \bibinfo{author}{\bibfnamefont{J.~M.} \bibnamefont{Vogels}},
  \bibnamefont{and} \bibinfo{author}{\bibfnamefont{W.}~\bibnamefont{Ketterle}},
  \bibinfo{journal}{Science} \textbf{\bibinfo{volume}{292}},
  \bibinfo{pages}{476} (\bibinfo{year}{2001}).

\bibitem[{\citenamefont{Shin et~al.}(2004)\citenamefont{Shin, Saba,
  Vengalattore, Pasquini, Sanner, Leanhardt, Prentiss, Pritchard, and
  Ketterle}}]{ketterle:dynamical_instability_doubly_quantized_vortex}
\bibinfo{author}{\bibfnamefont{Y.}~\bibnamefont{Shin}},
  \bibinfo{author}{\bibfnamefont{M.}~\bibnamefont{Saba}},
  \bibinfo{author}{\bibfnamefont{M.}~\bibnamefont{Vengalattore}},
  \bibinfo{author}{\bibfnamefont{T.~A.} \bibnamefont{Pasquini}},
  \bibinfo{author}{\bibfnamefont{C.}~\bibnamefont{Sanner}},
  \bibinfo{author}{\bibfnamefont{A.~E.} \bibnamefont{Leanhardt}},
  \bibinfo{author}{\bibfnamefont{M.}~\bibnamefont{Prentiss}},
  \bibinfo{author}{\bibfnamefont{D.~E.} \bibnamefont{Pritchard}},
  \bibnamefont{and} \bibinfo{author}{\bibfnamefont{W.}~\bibnamefont{Ketterle}},
  \bibinfo{journal}{Phys. Rev. Lett.} \textbf{\bibinfo{volume}{93}},
  \bibinfo{pages}{160406} (\bibinfo{year}{2004}).

\bibitem[{\citenamefont{Anderson et~al.}(2000)\citenamefont{Anderson, Haljan,
  Wieman, and Cornell}}]{cornell:vortex_precession_filled_and_empty_cores}
\bibinfo{author}{\bibfnamefont{B.~P.} \bibnamefont{Anderson}},
  \bibinfo{author}{\bibfnamefont{P.~C.} \bibnamefont{Haljan}},
  \bibinfo{author}{\bibfnamefont{C.~E.} \bibnamefont{Wieman}},
  \bibnamefont{and} \bibinfo{author}{\bibfnamefont{E.~A.}
  \bibnamefont{Cornell}}, \bibinfo{journal}{Phys. Rev. Lett.}
  \textbf{\bibinfo{volume}{85}}, \bibinfo{pages}{2857} (\bibinfo{year}{2000}).

\bibitem[{\citenamefont{Freilich et~al.}(2010)\citenamefont{Freilich, Bianchi,
  Kaufman, Langin, and Hall}}]{hall:dynamics_vortex_lines}
\bibinfo{author}{\bibfnamefont{D.~V.} \bibnamefont{Freilich}},
  \bibinfo{author}{\bibfnamefont{D.~M.} \bibnamefont{Bianchi}},
  \bibinfo{author}{\bibfnamefont{A.~M.} \bibnamefont{Kaufman}},
  \bibinfo{author}{\bibfnamefont{T.~K.} \bibnamefont{Langin}},
  \bibnamefont{and} \bibinfo{author}{\bibfnamefont{D.~S.} \bibnamefont{Hall}},
  \bibinfo{journal}{Science} \textbf{\bibinfo{volume}{329}},
  \bibinfo{pages}{1182} (\bibinfo{year}{2010}).

\bibitem[{\citenamefont{Schweikhard et~al.}(2004)\citenamefont{Schweikhard,
  Coddington, Engels, Mogendorff, and Cornell}}]{cornell:lowest_landau_level}
\bibinfo{author}{\bibfnamefont{V.}~\bibnamefont{Schweikhard}},
  \bibinfo{author}{\bibfnamefont{I.}~\bibnamefont{Coddington}},
  \bibinfo{author}{\bibfnamefont{P.}~\bibnamefont{Engels}},
  \bibinfo{author}{\bibfnamefont{V.~P.} \bibnamefont{Mogendorff}},
  \bibnamefont{and} \bibinfo{author}{\bibfnamefont{E.~A.}
  \bibnamefont{Cornell}}, \bibinfo{journal}{Phys. Rev. Lett.}
  \textbf{\bibinfo{volume}{92}}, \bibinfo{eid}{040404} (\bibinfo{year}{2004}).

\bibitem[{\citenamefont{Sorensen et~al.}(2005)\citenamefont{Sorensen, Demler,
  and Lukin}}]{sorensen:quantum_hall}
\bibinfo{author}{\bibfnamefont{A.}~\bibnamefont{Sorensen}},
  \bibinfo{author}{\bibfnamefont{E.}~\bibnamefont{Demler}}, \bibnamefont{and}
  \bibinfo{author}{\bibfnamefont{M.}~\bibnamefont{Lukin}},
  \bibinfo{journal}{Phys. Rev. Lett.} \textbf{\bibinfo{volume}{94}},
  \bibinfo{pages}{86803} (\bibinfo{year}{2005}).

\bibitem[{\citenamefont{Gemelke et~al.}(2010)\citenamefont{Gemelke, Sarajlic,
  and Chu}}]{2010arXiv1007}
\bibinfo{author}{\bibfnamefont{N.}~\bibnamefont{Gemelke}},
  \bibinfo{author}{\bibfnamefont{E.}~\bibnamefont{Sarajlic}}, \bibnamefont{and}
  \bibinfo{author}{\bibfnamefont{S.}~\bibnamefont{Chu}},
  \bibinfo{journal}{ArXiv e-prints}  (\bibinfo{year}{2010}),
  \eprint{1007.2677}.

\bibitem[{\citenamefont{Kosterlitz and Thouless}(1973)}]{kosterlitz:bkt}
\bibinfo{author}{\bibfnamefont{J.~M.} \bibnamefont{Kosterlitz}}
  \bibnamefont{and} \bibinfo{author}{\bibfnamefont{D.~J.}
  \bibnamefont{Thouless}}, \bibinfo{journal}{J. Phys. C}
  \textbf{\bibinfo{volume}{6}}, \bibinfo{pages}{1181} (\bibinfo{year}{1973}).

\bibitem[{\citenamefont{Hadzibabic et~al.}(2006)\citenamefont{Hadzibabic,
  Kr\"uger, Cheneau, Battelier, and Dalibard}}]{dalibard:2D_vortices}
\bibinfo{author}{\bibfnamefont{Z.}~\bibnamefont{Hadzibabic}},
  \bibinfo{author}{\bibfnamefont{P.}~\bibnamefont{Kr\"uger}},
  \bibinfo{author}{\bibfnamefont{M.}~\bibnamefont{Cheneau}},
  \bibinfo{author}{\bibfnamefont{B.}~\bibnamefont{Battelier}},
  \bibnamefont{and} \bibinfo{author}{\bibfnamefont{J.}~\bibnamefont{Dalibard}},
  \bibinfo{journal}{Nature} \textbf{\bibinfo{volume}{441}},
  \bibinfo{pages}{1118} (\bibinfo{year}{2006}).

\bibitem[{\citenamefont{Matthews
  et~al.}(1999{\natexlab{a}})\citenamefont{Matthews, Anderson, Haljan, Hall,
  Wieman, and Cornell}}]{cornell:vortices}
\bibinfo{author}{\bibfnamefont{M.~R.} \bibnamefont{Matthews}},
  \bibinfo{author}{\bibfnamefont{B.~P.} \bibnamefont{Anderson}},
  \bibinfo{author}{\bibfnamefont{P.~C.} \bibnamefont{Haljan}},
  \bibinfo{author}{\bibfnamefont{D.~S.} \bibnamefont{Hall}},
  \bibinfo{author}{\bibfnamefont{C.~E.} \bibnamefont{Wieman}},
  \bibnamefont{and} \bibinfo{author}{\bibfnamefont{E.~A.}
  \bibnamefont{Cornell}}, \bibinfo{journal}{Phys. Rev. Lett.}
  \textbf{\bibinfo{volume}{83}}, \bibinfo{pages}{2498}
  (\bibinfo{year}{1999}{\natexlab{a}}).

\bibitem[{\citenamefont{Matthews
  et~al.}(1999{\natexlab{b}})\citenamefont{Matthews, Anderson, Haljan, Hall,
  Wieman, and Cornell}}]{matthews:vortices}
\bibinfo{author}{\bibfnamefont{M.~R.} \bibnamefont{Matthews}},
  \bibinfo{author}{\bibfnamefont{B.~P.} \bibnamefont{Anderson}},
  \bibinfo{author}{\bibfnamefont{P.~C.} \bibnamefont{Haljan}},
  \bibinfo{author}{\bibfnamefont{D.~S.} \bibnamefont{Hall}},
  \bibinfo{author}{\bibfnamefont{C.~E.} \bibnamefont{Wieman}},
  \bibnamefont{and} \bibinfo{author}{\bibfnamefont{E.~A.}
  \bibnamefont{Cornell}}, \bibinfo{journal}{Phys. Rev. Lett.}
  \textbf{\bibinfo{volume}{83}}, \bibinfo{pages}{2498}
  (\bibinfo{year}{1999}{\natexlab{b}}).

\bibitem[{\citenamefont{Leanhardt et~al.}(2002)\citenamefont{Leanhardt,
  G\"orlitz, Chikkatur, Kielpinski, Shin, Pritchard, and
  Ketterle}}]{leanhardt:topological_vortices}
\bibinfo{author}{\bibfnamefont{A.~E.} \bibnamefont{Leanhardt}},
  \bibinfo{author}{\bibfnamefont{A.}~\bibnamefont{G\"orlitz}},
  \bibinfo{author}{\bibfnamefont{A.~P.} \bibnamefont{Chikkatur}},
  \bibinfo{author}{\bibfnamefont{D.}~\bibnamefont{Kielpinski}},
  \bibinfo{author}{\bibfnamefont{Y.}~\bibnamefont{Shin}},
  \bibinfo{author}{\bibfnamefont{D.~E.} \bibnamefont{Pritchard}},
  \bibnamefont{and} \bibinfo{author}{\bibfnamefont{W.}~\bibnamefont{Ketterle}},
  \bibinfo{journal}{Phys. Rev. Lett.} \textbf{\bibinfo{volume}{89}},
  \bibinfo{pages}{190403} (\bibinfo{year}{2002}).

\bibitem[{\citenamefont{Neely et~al.}(2010)\citenamefont{Neely, Samson,
  Bradley, Davis, and Anderson}}]{PhysRevLett.104.160401}
\bibinfo{author}{\bibfnamefont{T.~W.} \bibnamefont{Neely}},
  \bibinfo{author}{\bibfnamefont{E.~C.} \bibnamefont{Samson}},
  \bibinfo{author}{\bibfnamefont{A.~S.} \bibnamefont{Bradley}},
  \bibinfo{author}{\bibfnamefont{M.~J.} \bibnamefont{Davis}}, \bibnamefont{and}
  \bibinfo{author}{\bibfnamefont{B.~P.} \bibnamefont{Anderson}},
  \bibinfo{journal}{Phys. Rev. Lett.} \textbf{\bibinfo{volume}{104}},
  \bibinfo{pages}{160401} (\bibinfo{year}{2010}).

\bibitem[{\citenamefont{Turner}(2009)}]{PhysRevLett.103.080603}
\bibinfo{author}{\bibfnamefont{A.~M.} \bibnamefont{Turner}},
  \bibinfo{journal}{Phys. Rev. Lett.} \textbf{\bibinfo{volume}{103}},
  \bibinfo{pages}{080603} (\bibinfo{year}{2009}).

\bibitem[{\citenamefont{Khawaja and
  Stoof}(2001)}]{khawaja:skyrmions_in_ferromagnetic_bec}
\bibinfo{author}{\bibfnamefont{U.~A.} \bibnamefont{Khawaja}} \bibnamefont{and}
  \bibinfo{author}{\bibfnamefont{H.~T.~C.} \bibnamefont{Stoof}},
  \bibinfo{journal}{Nature} \textbf{\bibinfo{volume}{411}},
  \bibinfo{pages}{918} (\bibinfo{year}{2001}).

\bibitem[{\citenamefont{Skyrme}(1961)}]{skyrme:nonlinear_field_theory}
\bibinfo{author}{\bibfnamefont{T.~H.~R.} \bibnamefont{Skyrme}},
  \bibinfo{journal}{Proc. R. Soc. Lond. A} \textbf{\bibinfo{volume}{260}},
  \bibinfo{pages}{127} (\bibinfo{year}{1961}).

\bibitem[{\citenamefont{Leanhardt et~al.}(2003)\citenamefont{Leanhardt, Shin,
  Kielpinski, Pritchard, and Ketterle}}]{ketterle:coreless_vortex_bec}
\bibinfo{author}{\bibfnamefont{A.~E.} \bibnamefont{Leanhardt}},
  \bibinfo{author}{\bibfnamefont{Y.}~\bibnamefont{Shin}},
  \bibinfo{author}{\bibfnamefont{D.}~\bibnamefont{Kielpinski}},
  \bibinfo{author}{\bibfnamefont{D.~E.} \bibnamefont{Pritchard}},
  \bibnamefont{and} \bibinfo{author}{\bibfnamefont{W.}~\bibnamefont{Ketterle}},
  \bibinfo{journal}{Phys. Rev. Lett.} \textbf{\bibinfo{volume}{90}},
  \bibinfo{pages}{140403} (\bibinfo{year}{2003}).

\bibitem[{\citenamefont{Andrews et~al.}(1997)\citenamefont{Andrews, Townsend, ,
  Miesner, Durfee, Kurn, and Ketterle}}]{ketterle:interference_2_becs}
\bibinfo{author}{\bibfnamefont{M.~R.} \bibnamefont{Andrews}},
  \bibinfo{author}{\bibfnamefont{C.~G.} \bibnamefont{Townsend}}, ,
  \bibinfo{author}{\bibfnamefont{H.-J.} \bibnamefont{Miesner}},
  \bibinfo{author}{\bibfnamefont{D.~S.} \bibnamefont{Durfee}},
  \bibinfo{author}{\bibfnamefont{D.~M.} \bibnamefont{Kurn}}, \bibnamefont{and}
  \bibinfo{author}{\bibfnamefont{W.}~\bibnamefont{Ketterle}},
  \bibinfo{journal}{Science} \textbf{\bibinfo{volume}{275}},
  \bibinfo{pages}{637} (\bibinfo{year}{1997}).

\bibitem[{\citenamefont{Bakr et~al.}(2009)\citenamefont{Bakr, Gillen, Peng,
  F\"olling, and Greiner}}]{greiner:quantum_gas_microscope}
\bibinfo{author}{\bibfnamefont{W.~S.} \bibnamefont{Bakr}},
  \bibinfo{author}{\bibfnamefont{J.~I.} \bibnamefont{Gillen}},
  \bibinfo{author}{\bibfnamefont{A.}~\bibnamefont{Peng}},
  \bibinfo{author}{\bibfnamefont{S.}~\bibnamefont{F\"olling}},
  \bibnamefont{and} \bibinfo{author}{\bibfnamefont{M.}~\bibnamefont{Greiner}},
  \bibinfo{journal}{Nature} \textbf{\bibinfo{volume}{462}}, \bibinfo{pages}{74}
  (\bibinfo{year}{2009}).

\end{thebibliography}
\end{document}